\documentclass[12pt,preprint]{aastex}

\begin{document}

\title{Viewpoints: A high-performance high-dimensional exploratory
data analysis tool}

\author{P.R. Gazis}
\affil{SETI Institute, Mountain View, CA, 94043}
\email{pgazis@sbcglobal.net}

\author{C. Levit}
\affil{NASA Ames Research Center, Moffett Field, CA 94035}
\email{Creon.Levit@nasa.gov}

\and

\author{M.J. Way\altaffilmark{1,2}}
\affil{NASA Goddard Institute for Space Studies,
2880 Broadway, New York, NY, 10025}
\email{Michael.J.Way@nasa.gov}

\altaffiltext{1}{NASA Ames Research Center, Space Science Division,
    Moffett Field, CA 94035, USA}
\altaffiltext{2}{Department of Astronomy and Space Physics,
Uppsala, Sweden}

\begin{abstract}
Scientific data sets continue to increase in both size and complexity.  In 
the past, dedicated graphics systems at supercomputing centers were required 
to visualize large data sets, but as the price of commodity graphics hardware
has dropped and its capability has increased, it is now possible, in principle,
to view large complex data sets on a single workstation. To do this in practice,
an investigator will need software that is written to take advantage of the 
relevant graphics hardware.  The Viewpoints visualization package described 
herein is an example of such software.
Viewpoints is an interactive tool for exploratory visual analysis of large, 
high-dimensional (multivariate) data.  It leverages the capabilities of 
modern graphics boards (GPUs) to run on a single workstation or laptop.  
Viewpoints is minimalist:  it attempts to do a small set of useful things 
very well (or at least very quickly) in comparison with similar packages today.
Its basic feature set includes linked scatter plots with brushing, dynamic
histograms, normalization and outlier detection/removal.
Viewpoints was originally designed for 
astrophysicists, but it has since been used in a variety of fields that
range from astronomy, quantum chemistry, fluid dynamics, machine learning, 
bioinformatics, and finance to information technology server log mining.
In this article, we describe the Viewpoints package and show examples
of its usage.

\end{abstract}


\section{Introduction}

Analysis and visualization of extremely large and complex data sets has 
become one of the more significant challenges facing the scientific 
community today. Current instruments and simulations produce enormous 
volumes of data. These data sets include hyperspectral images from 
spacecraft, multivariate data of high dimensionality from sky surveys, 
time-varying three-dimensional flows from supercomputer simulations, and 
complex interrelated time series from vehicle telemetry, DNA microarrays 
or physiological monitoring, to name a few.  

One representative example from astronomy is the Sloan Digital Sky 
\citep[SDSS][]{York2000}. The spectroscopic catalog from the SDSS Data 
Release 7 \citep[DR7][]{Abazajian2009} contains information for over 1.5
million objects.  For each object in the joint spectroscopic and photometric
catalog there are over 100 physical measurements.  Properties in these catalogs
include position on the sky, spectroscopic redshift, photometric magnitude
in five bands, energy/width of up to 50 spectral lines, galaxy angular size,
bulge-to-disk ratio, etc. In addition, many other physical quantities of
interest can be derived from these data, such as absolute magnitudes,
photometric redshifts, galaxy type, average distance to nearest neighbors
and other properties of the galaxy's local environment.  This is the kind
of catalog in which Viewpoints excels on a relatively modern desktop computer.

On the other hand the DR7 photometric catalog has measured over 350 million
unique objects. This is an enormous amount of data by traditional standards
and these kinds of data volumes will continue to grow in size over the
coming decade.  The actual data volume in the SDSS DR7, including pre- and 
postprocessing products, is more than 50 TB, while the searchable database for 
the photometric and redshift catalogs is over 
3.5 TB.\footnote{http://www.sdss.org/dr7}
We do not claim that Viewpoints is able to explore a data set of this
size all at once as it is beyond the memory capabilities of a normal
desktop computer. Still, Viewpoints can quickly explore subsets of
a large complex data set like this. When Viewpoints does reach the
limits of desktop hardware, one can reach for more capable, but expensive
and rare, custom-built hardware systems such as the
hyperwall \citep{Sandstrom2003}. The hyperwall can then be utilized
efficiently using the knowledge already obtained through exploration
via Viewpoints.

Although advances in hardware speed and storage technology have kept up with 
(and, indeed, have driven) increases in database size, the same is not true 
of the tools we use to explore and understand these data
\citep[e.g.,][]{Subbarao08}.
Modern data sets routinely outstrip the capabilities of contemporary
visualization and analysis software, and this problem can only get worse as
data volumes from future astronomy-related programs
(e.g., Large Synoptic Survey Telescope [LSST]\footnote{See http://www.lsst.org},
Supernova Acceleration Probe [SNAP]\footnote{See http://snap.lbl.gov},
Panoramic Survey Telescope and Rapid Response System
[PanSTARRS]\footnote{See http://pan-starrs.ifa.hawaii.edu/public}, and
Dark Energy Survey [DES]\footnote{See http://www.darkenergysurvey.org})
and ever-larger supercomputer simulations increase by orders of magnitude. 
See \cite{Borne2009} for a review.
Data sets of this size and dimensionality defy interactive analysis with 
standard tools such as MATLAB,\footnote{See http://www.mathworks.com}
Octave,\footnote{See http://www.octave.org}
Mathematica,\footnote{See http://www.wolfram.com}
R,\footnote{See http://www.r-project.org}
S++,\footnote{See http://www.splusplus.com}
xGobi,\footnote{See http://www.research.att.com/areas/stat/xgobi}
Mirage,\footnote{See http://www.bell-labs.com/project/mirage}
gnuplot,\footnote{See http://www.gnuplot.info} etc.
To address this problem, we have developed Viewpoints.

Viewpoints was inspired to a large extent by the hyperwall 
\citep{Sandstrom2003}. This facility, developed by the 
NASA Advanced Supercomputing division at NASA Ames Research Center,
is a high-performance visualization cluster that was specifically designed 
to address the problem of exploring, visualizing, and analyzing large, 
complex, multidimensional data. See \cite{Wong1997} for a review on
the history of multidimensional multivariate visualization.  In its original
form, the hyperwall was a 50 node Linux cluster with a 7$\times$7 array of
tiled displays that could be used to display data in a variety of formats that 
might range from simple plots to elaborate 3D rendering.  It has since been 
superseded by even larger facilities at NASA Ames Research Center and 
several other institutions throughout the world.  

Hyperwalls have been applied to a wide range of problems in astrophysics, 
Earth science, aerodynamics, and life sciences
\citep{Macdoughall2003,Brown2004,Murman2004}. 
Unfortunately, tools such as the hyperwall require custom hardware and a 
team of onsite specialists at a dedicated facility, which places them 
beyond the reach of most investigators and research programs.  

\begin{figure}[!htb]
\plotone{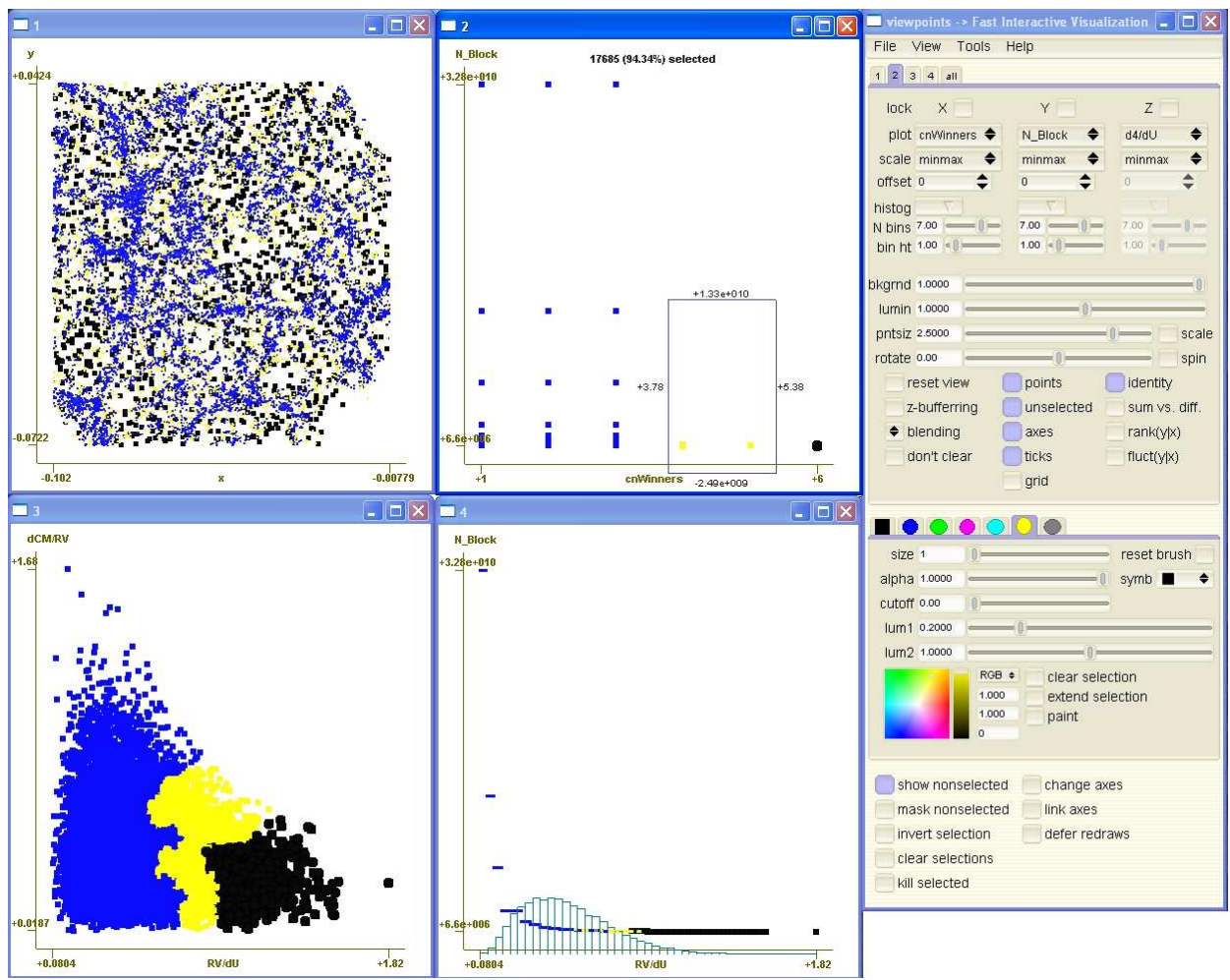}
\caption{Viewpoints displaying a 2$\times$2 array of scatter plots of SDSS data.
(Upper left) xy positions of $\sim$20,000 galaxies from the SDSS.  Colors indicate
different classes identified by a clustering algorithm.  
(Upper right) Density classes identified by a Bayesian partitioning 
scheme vs class identified by a Self-Organizing Map.
(Lower left) Local density vs local gradient.
(Lower right)  Local density vs density classes identified by a Bayesian 
partitioning scheme. 
(Far right) The viewpoints control panel.}
\label{fig:sdss_figure_001}
\end{figure}

Viewpoints, shown in Figure~\ref{fig:sdss_figure_001}, is intended to provide a 
small but powerful subset of the hyperwall's functionality on any working 
scientist's desktop or laptop computer.  
Viewpoints provides a high-performance interactive multidimensional visual
exploratory data analysis environment that allows users to quickly explore
and visually analyze the most common forms of large multidimensional
scientific data. It can also be run as a ``plug in" or external module for
scientific programming environments such as MATLAB.

In contrast to programs like Mirage \citep{Ho2007} or
VisIVO \citep{Becciani2010} which have a large number of supporting features,
Viewpoints does a few things very well.
Its basic feature set includes linked scatter plots with brushing
\citep[e.g.,][]{BC1987,DS2000,HCHEN2003,Stump2004,Co2005,Jordan2008},
dynamic histograms, normalization, and outlier detection/removal.  It also
includes a limited capability for Boolean operations with brushes analogous 
to those of \cite{Hurter2009}.

\section{Overview and Capabilities}

Viewpoints is a fast interactive tool for the analysis and visualization of
high-dimensional data primarily via linked scatter plots with brushing.  
It is intended for general-purpose visual data analysis, interactive 
exploratory data mining, and machine-assisted pattern discovery.  The 
targeted application is the exploration of large, complex, multivariate data 
sets, with dimensionalities up to 100 or more and sample sizes up to
10$^{5}$--10$^{7}$.  
Viewpoints takes advantage of hardware-accelerated graphics
(graphics processing units [GPUs]) 
and other capabilities of modern desktop workstations and laptops  
(e.g., multithreading and memory-mapped input/output).  
As previously above, it is highly interactive, and almost all of its parameters 
can be modified in real time to provide immediate visual feedback.

Viewpoints can be invoked either from a desktop icon or from the command 
line with a set of optional parameters. 
Once it is running, it is controlled by a conventional GUI 
and/or configuration files saved from previous sessions.
Viewpoints reads multidimensional data in the form of ASCII files, binary 
files, or simple Flexible Image Transport System (FITS) tables
and displays these data as an arbitrary number
of simultaneously linked 2D and 3D scatter plots 
\citep[e.g.,][]{Comparato2007} with independently 
controllable viewing transformations, normalization schemes, overplotting 
schemes \citep{Zhang2003}, axis labels, and other formatting. These data can
then be visualized and explored interactively using a set of graphical, 
statistical, and semiautomated pattern-recognition techniques.  The user 
can select, flag, and/or delete subsets of the data using brushes, and the 
modified data set can be written back to disk in ASCII, binary, or FITS 
format.

The program's capabilities were carefully selected using a minimalist 
philosophy.  We have chosen to provide a small core of essential features 
that are universally useful and keep the program size small, execution 
speeds high, and the learning curves shallow.
Figure~\ref{fig:sdss_figure_001} shows an example of Viewpoints displaying
a small 2$\times$2 array of linked scatter plots.  This figure is described in
detail in \S \ref{sec:Examples}.  In practice, one can 
simultaneously display an arbitrary number of linked scatter plots, and we 
routinely use up to 30 windows to examine data sets of high dimensionality.
Viewpoints has a conventional GUI, shown in the
right panel of Figure~\ref{fig:sdss_figure_001}, with an assortment of menus,
buttons, and 
sliders that provide access to its features.  The most important of these
capabilities can also be invoked by mouse gestures and/or keyboard 
shortcuts.  It has a full range of basic help facilities, such as usage 
messages, help files, and tooltips.  It also produces a diagnostic output 
stream when invoked via the command line.

Viewpoints has a tightly integrated set of features.
A partial list of these features is provided next. Some were in the original
release, others were added in response to users requests, and a few were
contributed to the open source
repository\footnote{See http://www.assembla.com/wiki/show/viewpoints}
by a small community of developers.

\begin{enumerate}
\item{Multiple linked scatter plots with dynamic brushing rendered via OpenGL 
with update rates of greater than $10^7$ points $s^{-1}$ for most modern
graphics cards.}
\item{The ability to brush separate selections using separate colors and/or 
symbols.}
\item{The ability to map densely overplotted regions smoothly to different 
hues and/or intensities.}
\item{The ability to save the program's complete state to an XML formatted
configuration file at any time.}
\item{The ability to initialize and/or restore the program's state from a
configuration file.}
\item{The ability to append and merge additional data and create or destroy
additional plot windows on the fly.}
\item{Per-axis one-dimensional equiwidth marginal histograms with
dynamically adjustable bin widths and dynamic update of the selected 
fraction in each bin.  Histograms are individually selectable and adjustable 
for both the $x$ and $y$ axes of each plot separately.}
\item{Per-axis automatic normalizations based data limits, quantiles, 
moments, rank, or simple Gaussianization.}
\item{Elementary outlier detection and removal.}
\item{The ability to change most parameters globally across all plots
simultaneously, or locally on a single plot.}
\item{The ability to lock (and unlock) individual plots so they are immune
from (susceptible to) global operations.}
\item{Rudimentary semi-automated browsing through high-dimensional data by
permuting all (unlocked) axes on all plots.}
\item{Real-time display of quantitative data for individual axes and for 
the current selection.}
\item{Basic online help.}
\item{A comprehensive set of error tests along with graceful error recovery
if a user attempts to read corrupt and ill-formatted data or configuration
files.}
\item{The ability to replace missing values with user-specified default 
values.}
\item{Basic ability to plot time series analogous to that
of \cite{AkibaMa2007}.}
\item{Viewpoints also achieves smooth interactive performance on large data as
long as:
\begin{enumerate}
\item The (binary) array corresponding to the dataset fits into real
system (CPU) memory 
\item The coordinates of all vertices being displayed fit into available
graphics (GPU) memory. For example, on a 512 Mbyte graphics card viewpoints
can simultaneously update and display
(4 plots)$\times$(8x10$^{6}$points/plot)$\times$(3 vertices/point)$\times$
(4 bytes/vertex)= $\sim$400Megabytes
\end{enumerate}}
\end{enumerate}

\section{Software Design}

Viewpoints uses hardware-accelerated OpenGL 1.5 on all platforms and 
performs rendering via OpenGL\footnote{http://www.opengl.org} vertex buffer 
objects to provide extremely high graphics performance.  The package has 
cross-platform availability and can be easily ported to other architectures. 
Table \ref{tbl-1} has a list of operating systems and architectures known
to work to-date. Viewpoints has a preliminary interface to MATLAB and Octave.
It has the potential to work with scripting languages such as 
Python\footnote{See http://www.python.org} or 
Perl.\footnote{See http://www.perl.org} Most importantly, Viewpoints is
specifically engineered for use with large data sets.  
As noted previously, the targeted application is the interactive analysis of 
large, complex, multivariate data sets, with dimensionalities that may 
surpass 100 and sample sizes that may exceed 10$^{5}$--10$^{7}$.

\begin{deluxetable}{lll}
\tablecolumns{3}
\tablewidth{0pc}
\tablecaption{Platforms Known To Run viewpoints\label{tbl-1}}
\tablehead{
\colhead{Operating System} &
\colhead{OS Revision} &
\colhead{Architecture}
}
\startdata
Apple     & Mac OS 10.3--10.6                            & Intel \\
Apple     & Mac OS 10.3--10.6                            & PPC   \\
Linux     & Most Distributions with Kernels 2.4.x--2.6.x & Intel \\
Windows   & 2000                                         & Intel \\
Windows   & XP                                           & Intel \\
Windows   & Vista                                        & Intel \\
Windows   & 7                                            & Intel \\
\enddata
\end{deluxetable}

In the interests of portability, extensibility and efficiency, Viewpoints 
is written in C++.  This allows us to program at a high level of abstraction 
and still achieve excellent runtime performance on all target platforms.  
Most importantly, C++ allows us to leverage vast amounts of code written by 
other people.  Identical code compiles and runs on Linux, Microsoft Windows, 
and Apple OS X.

Viewpoints is a relatively small program. The compiled Linux binary is only 
5.5 Mbytes.\footnote{The compiled file contains a
large number of statically linked libraries} Since it makes extensive use of 
libraries including some that leverage C++ template metaprogramming
(e.g., Blitz++\footnote{http://www.oonumerics.org/blitz} and 
Boost\footnote{http://www.boost.org}), the Viewpoints source tree
itself is also rather small (approximately 40,000 lines).
The libraries it depends on are listed in Table \ref{tbl-2}.

\begin{deluxetable}{ll}
\tablecolumns{2}
\rotate
\tablewidth{0pc}
\tablecaption{C++ libraries used in viewpoints\label{tbl-2}}
\tabletypesize{\scriptsize}
\tablehead{
\colhead{Library} &
\colhead{Description}}
\startdata
libc++    & The standard runtime library for basic math, input/output, and
memory management\\
STL       & The standard template library for data structures and algorithms\\
OpenGL    & The standard cross-platform library for hardware
accelerated graphics\\
FLTK      & A lightweight platform-independent open source graphical user
interface (GUI) library\\
Blitz++   & A library of extensions that add FORTRAN 90-style array
expressions (and fortran-90 style numerical performance) to c++\\
GSL       & The GNU scientific library. Includes functionality such as
statistics and linear algebra\\
Boost serialization & a library of extensions for saving and restoring the state of c++ objects.
\enddata
\end{deluxetable}

\section{GPU Usage and OpenGL Implementation}

Graphics boards that were available at the time development began, such as 
NVIDIA Geforce 7950, can transform and display over a billion floating-point 
vertices per second and can drive two or more 4 Mpixel displays
simultaneously. This capability continues to increase.  In its default mode,
Viewpoints uses GPUs explicitly to take advantage of this tremendous 
processing and rendering power.  This allows the package to provide smooth
high-frame-rate interactivity while manipulating multiple linked views, each
containing millions of data points.  
If for some reason a GPU is not available, Viewpoints can also be instructed 
to use a system's ordinary CPU, at some cost in performance.

The current release uses the following advanced OpenGL features:

\begin{enumerate}
\item{GPU-resident vertex buffer objects (VBOs) for point coordinates.
This conserves main memory (by offloading relevant data to GPU memory),
increases graphics performance (by caching frequently rendered vertex data
locally on the GPU), and frees up essentially all of the CPU bandwidth and 
bus bandwidth for other uses.} 
\item{GPU-resident VBOs for point indices. A single copy of the vertex
index arrays (which encode the selection status) is shared by all plots.}
\item{Squared alpha blending with separate blending control per plot. For data 
with wide variations in projected density this provides fast, adjustable 
mapping of overplot density to varying hue and/or intensity.}
\item{Point sprites for displaying symbols and/or smooth kernels instead
of just simple points (splatting).}  
\item{Antialiased points and/or point sprites of variable color, brightness, 
and size per brush and/or per plot.}
\item{Floating-point color and alpha values, allowing both more sensitivity
and more dynamic range for overplotting.}
\end{enumerate}

These design features provide dramatic graphics performance.  When one of 
our early beta releases was run on a 1997 MacBook Pro laptop, we 
measured throughput of over $70 \times 10^{6}$  vertices s$^{-1}$.
This consisted of six linked plots, each with 2 million data points
simultaneously updating at five frames s$^{-1}$.

Viewpoints has proved quite useful on small-sized (e.g., netbooks)
and medium-sized computer platforms with as little
as 1Gbyte of RAM. We do most of our demos on
a laptop displaying up to 18-27 variables in 6-9 windows on a single monitor.
To take full advantage of this package we recommend using a high-end
workstation with high-end graphics card, two high-resolution displays, and
several gigabytes of RAM. Experience suggests that this is a good 
impedance match to the visual perception of a scientist or a team of up 
to three collaborators.

The best way to think about the limitations of Viewpoints with respect
to common desktop computer hardware is that the data should be able
to literally fit in RAM memory. Specifically Viewpoints can achieve
smooth interactive performance on data as long as:
\begin{enumerate}
\item The (binary) array corresponding to the data set fits into real
system (CPU) memory
\item The coordinates of all vertices being displayed fit into available
graphics (GPU) memory.
\end{enumerate}

For those looking to the future please note that the relationship between the
amount of memory on a graphics card and the allowable size of the data set
appears to be linear. So the latest graphics cards with 1.5Gbyte GPU memory
should display four plots with $\sim$25$\times10^{6}$ points per plot, with an
update rate of a few frames per second.  The update rate is harder to estimate,
since ``vertices per second'' is no longer a common figure of merit in GPU specs.
However, this appears to scale roughly along with clock speed and GPU memory.

When visible plots exhaust GPU memory, one can run Viewpoints with the
``--no\_vbo'' flag, and check the ``defer redraws'' box in the lower right-hand
control panel. We have run 100 million point data sets on a laptop
this way and it is quite usable.

\section{Examples}\label{sec:Examples}

Figure~\ref{fig:sdss_figure_001} shows a simple example of Viewpoints usage with
a small 2$\times$2 array of linked scatter plots of data from the SDSS
\citep{York2000}, several methods of calculating local density, and a cluster 
detection scheme.  The upper left panel shows $xy$ coordinates of $\sim$20,000
galaxies from the SDSS.  When used this way, a scatter plot becomes
a map of a region of space that is approximately 30 Mpc across.
The blue, yellow, and black colors indicate 
cluster galaxies, halo galaxies, and field galaxies identified by a 
partitioning scheme based on self-organizing maps (SOMs).  It can be seen 
that cluster galaxies (blue) are concentrated into filaments while the field 
galaxies (black) are distributed more or less uniformly.
The upper right panel shows density classes identified by a Bayesian 
partitioning scheme versus classes identified by the SOM-based algorithm.
This panel is where the selection capability of 
Viewpoints was used.  The rectangle in the lower right side of the panel is 
the selection box that was used to flag the halo galaxies (yellow points).
The lower left panel shows local density versus local gradient to display the
distribution of the three galaxy classes in parameter space.
The lower right panel shows local density versus density class identified by a 
Bayesian partitioning scheme.  This panel shows a simple example of 
the histogram capability of Viewpoints. Figure~\ref{fig:sdss_figure_001}
provides a real-world example of the abilities of Viewpoints to quickly
deduce scientifically useful features for research purposes. The example
provided here is directly related to a recent publication by two of
the authors of this article \citep{WGS2010}.
All of the figures comparing the three different clustering technologies 
described in \citep{WGS2010} were originally devised
using Viewpoints. Publication-quality plots were later produced using MATLAB.

\begin{figure}[!htb]
\plotone{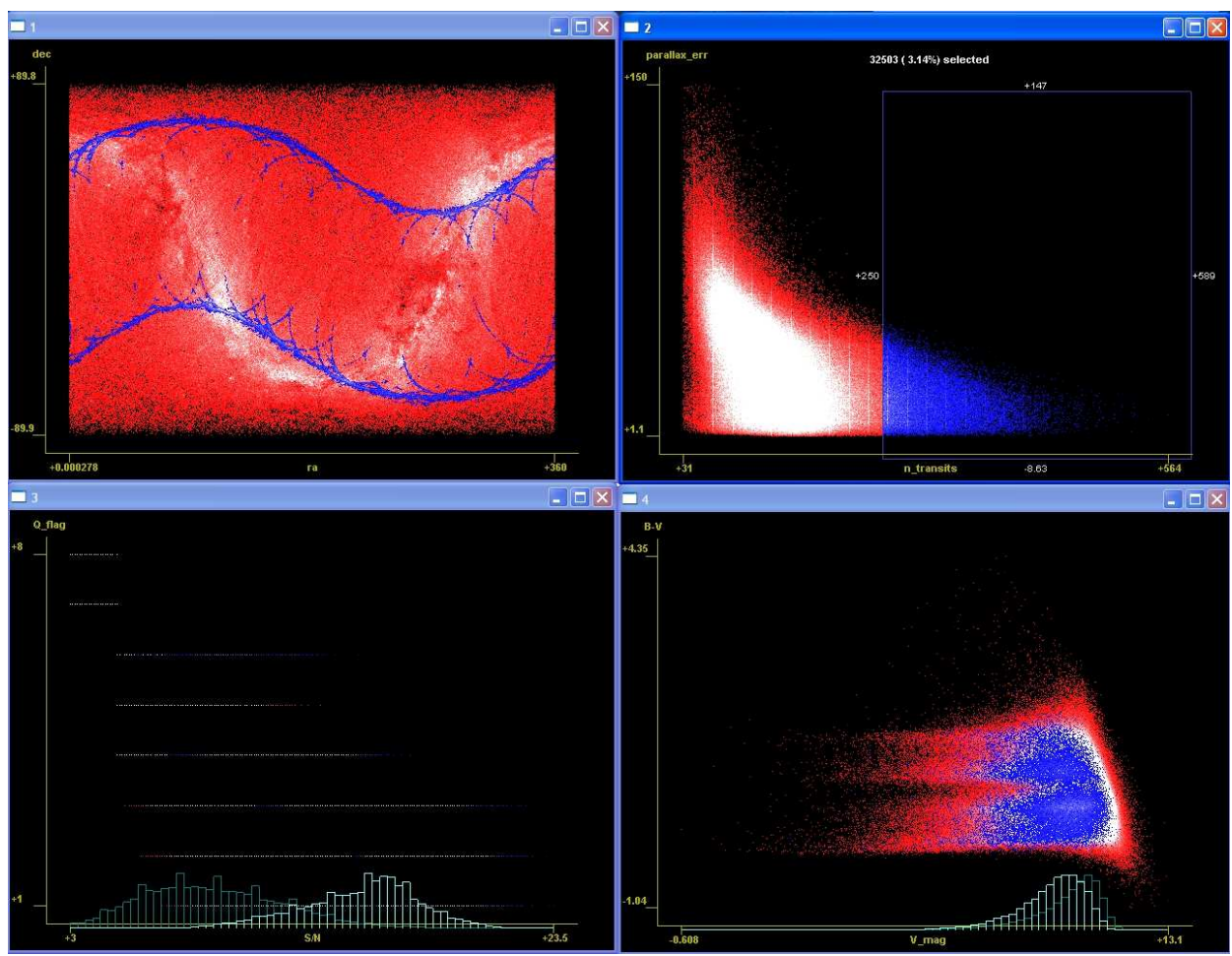}
\caption{Viewpoints displaying a 2$\times$2 array of linked scatter plots of 
Tycho data for 1 million stars.
(Upper left) Right ascension vs declination.
(Upper right) Number of transits vs parallax error, with a selection box
outlined in white.
(Lower left) Signal to noise ratio vs a quality flag.
(Lower right) V band magnitude plotted vs the B-V magnitude difference.}
\label{tycho_figure_002}
\end{figure}

Figure~\ref{tycho_figure_002} shows a more elaborate usage involving a
2$\times$2
array of linked scatter plots of astrometric data from the Tycho mission
\citep{ESA1997,Hog2000}.
For clarity, the control panel has been eliminated from this figure.
The upper left panel shows a plot of right ascension versus declination for 1 
million stars.  This panel illustrates the effect of overplotting with alpha
blending.  The galactic plane, with its greater concentration of stars, is 
plainly visible as a curving white band that meanders through the image.  
The upper right panel shows a plot of the number of transits versus parallax 
error.  The selection box (outlined in blue) has been used to select points 
for which more than 250 transits were available.  The legend at the top of 
the panel reports that these involved 32,503 points or 3.14\% of the total.  
This selection is repeated in the other panels.  In the upper left panel 
described previously, it appears as two parallel and roughly sinusoidal 
bands that mark the part of the spacecraft's orbit from which it was 
possible to obtain this number of transits.  The lower
left panel shows a plot of signal-to-noise ratio (S/N) versus a quality 
flag.  This is included to illustrate use of a marginal (conventional)
and a conditional histogram.  The latter peaks at a significantly higher
S/N ratio than the former, as would be expected, since 
measurements that involve more transits should have a higher S/N.
In the lower right panel the $V$-band magnitude has been plotted against
the $B--V$ magnitude difference to produce a typical color-magnitude diagram.

\begin{figure}[!htb]
\plotone{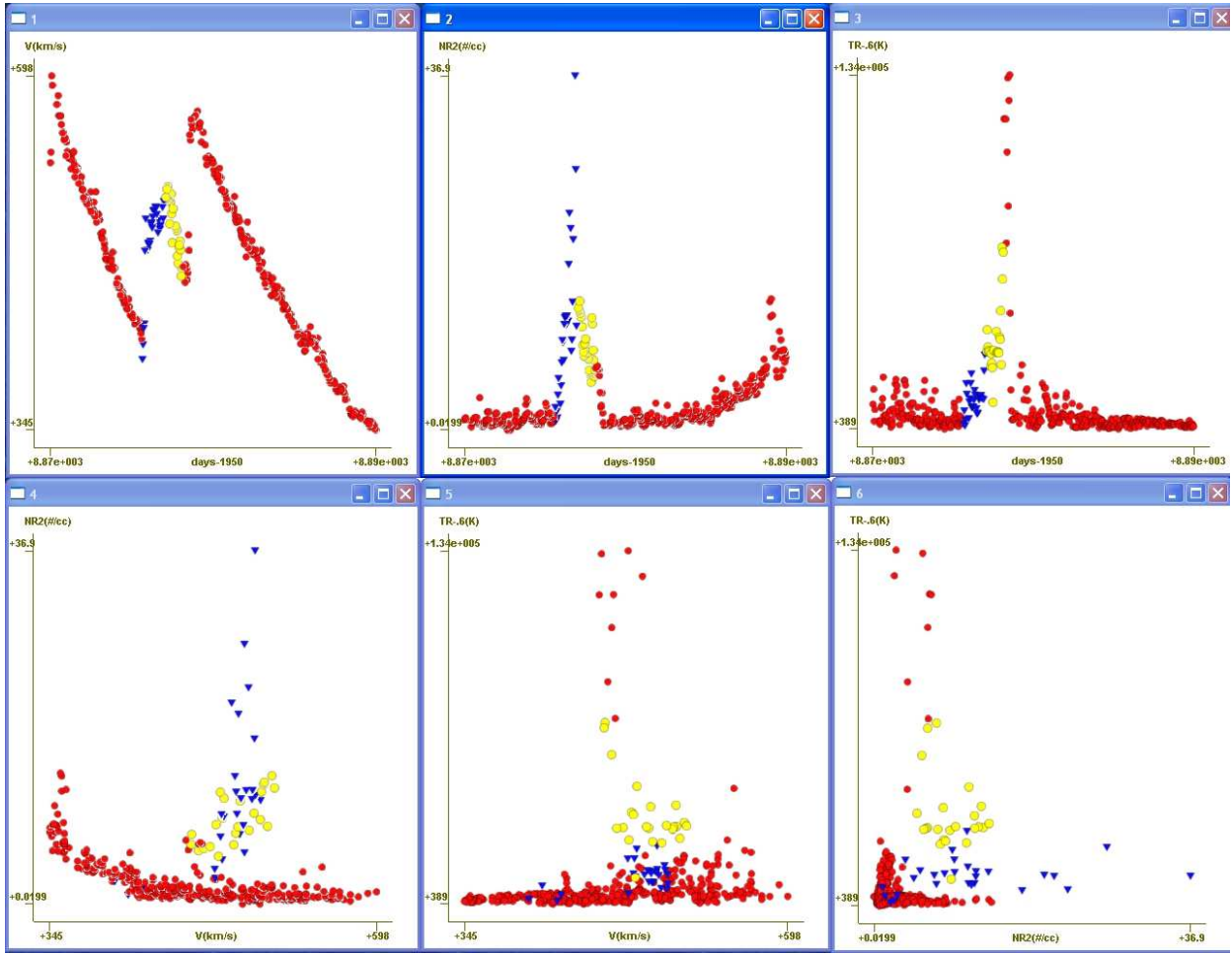}
\caption{Viewpoints displaying a 2$\times$3 array of linked scatter plots
of solar wind data.
(Upper left) Solar wind speed versus time.
(Upper middle) Solar wind density vs time.
(Upper right) Solar wind temperature vs time.
(Lower left) Density vs speed.
(Lower middle) Temperature vs speed.
(Lower right) Temperature vs density.}
\label{fig:sw_figure_003}
\end{figure}

Figure~\ref{fig:sw_figure_003} shows an example of Viewpoints used to visualize
the set of solar wind data that is provided with the Viewpoints package for 
demonstration purposes.  
In the top three panels, solar wind speed, density, and temperature have 
been plotted versus day since 1950 to produce a conventional time series.  These
show a typical structure in the solar wind, in which high-speed/low-density 
plasma (visible on the left of each panel) overtakes low-speed/low-density 
material (visible on the right) to produce an interaction region (visible in the
middle) that is characterized by intermediate speeds, high density, and 
high temperature.
These interaction regions are often subdivided into a leading portion and
trailing portion with different ranges of densities and temperatures
separated by a well-defined boundary. 
These are plainly visible in this time-series plot.
The selection capability of Viewpoints has been used to select these regions 
as blue and yellow.
The bottom three panels show scatter plots of velocity versus density, velocity
versus temperature, and density versus temperature.
It can be seen that the leading
portion of this interaction region (blue) was associated with a large range 
of densities and moderate range of temperatures, while the trailing portion
was associated with a moderate range 
of densities and large range of temperatures.

As mentioned in the Abstract, Viewpoints has been used extensively in a variety
of fields to advance specific scientific goals. Further examples include:
\begin{itemize}
\item Geophysics, Relationships between earthquake occurrence frequency,
location, season, and time of day.

\item Information Technology, Relationships between query response time,
query source location, time of day, type of query, etc. These were causing
slow response and high system loads on commercial servers which viewpoints
helped resolve quickly.

\item Finance, Currency arbitrage relationships and portfolio clustering
in statistical arbitrage.

\end{itemize}

\section{Opening up to the open source community}

Two significant concerns with any software tools are long-term support and 
response to the user community. Viewpoints has been released as an open source 
and is on a collaborative-oriented public wiki site called 
Assembla.\footnote{See http://www.assembla.com}
We now have a significant and growing user community that is successfully 
and enthusiastically applying Viewpoints to a huge variety of problem spread 
across multiple disciplines. 

\section{Conclusions and the future of viewpoints}

Viewpoints was adapted from its alpha prototype to general use with seed 
funding from the NASA Applied Information Systems Research
Program.\footnote{See http://aisrp.nasa.gov} 
It is now available on all major computing platforms (see Table 
\ref{tbl-1}) and is in use by scientists and engineers across all NASA
mission directorates, as well as by some early adopters in industry and
academia.  Its operation, interface, source code, and impressive performance
are the same on all platforms.  Viewpoints is now directly callable from 
MATLAB.  There is a Viewpoints wiki and
Web site\footnote{See http://www.assembla.com/wiki/show/viewpoints} that allows 
the open source community to take part in development and affords users 
access to the latest information regarding bug fixes and program 
enhancements.  In the near future we expect more development to come from 
the open source community and plan to modify the base code to take 
advantages of new advances in video chip technology as it continues to evolve.

\acknowledgments

Many thanks go to Chris Henze and Jeff Scargle for their input and
ideas. The authors would also like to thank Joe Bredekamp and his
NASA Applied Information Systems Research Program for pilot funding of
Viewpoints. M.J.W. would like to thank the Astronomy Department of Uppsala
University for their kind hospitality. This research has made use
of NASA's Astrophysics Data System.

Viewpoints has been released under the NASA Open Source Agreement and
is considered an open source license. The source code and precompiled binaries
can be freely downloaded from the Viewpoints Web site.
If you use Viewpoints in your research, please consider referencing this work.


\begin{thebibliography}{}

\bibitem[Abazajian et al.(2009)]{Abazajian2009}
Abazajian, K. et al. 2009, \apjs, 182, 543

\bibitem[Akiba \& Ma(2007)]{AkibaMa2007}
Akiba, H. \& Kwan-Liu, M. 2007, ``An Interactive Interface for Visualizing
Time-Varying Multivariate Volume Data", Proceedings of the Joint
Eurographics-IEEE VGTC Symposium on Visualization, May 2007

\bibitem[Becciani et al.(2010)]{Becciani2010}
Becciani, U. et al. 2010, \pasp, 122, 119

\bibitem[Becker \& Cleveland(1987)]{BC1987}
Becker, R.A. \& Cleveland, W.S. 1987, Technometrics, 29, 127

\bibitem[Borne(2009)]{Borne2009}
Born, K. 2009, arXiv:0911.0505v1

\bibitem[Brown et al.(2004)]{Brown2004}
Brown, J., Tobak, M., Prabhu, D. \& Sandstrom, T. ``Flow Topology About an
Orbiter Leading Edge Cavity at STS-107 Reentry Conditions'',
AIAA-2004-2285 37th AIAA Thermophysics Conference, Portland, Oregon, 2004.

\bibitem[Chen(2003)]{HCHEN2003}
Chen, H. 2003, IEEE Symp. on Inf. Vis. (Washington: IEEE), 181

\bibitem[Co et al.(2005)]{Co2005}
Co, C.S. et al. 2005, Eurographics/IEEE-VGTC Symp. Vis.
(Washington: IEEE) 279

\bibitem[Comparato et al.(2007)]{Comparato2007}
Comparato, M. et al. 2007, \pasp, 119, 898

\bibitem[Davidson \& Sardy(2000)]{DS2000}
Davidson, A.C. \& Sardy, S. 2000, J. Comput. and Graph. Stat., 9, 750

\bibitem[Ho(2007)]{Ho2007}
Ho, T.K. 2007, in Statistical Challenges in Modern Astronomy IV, ASP
Conferences Series, Vol. 371, p.391. eds G. Jogesh Babu and Eric D. Feigelson

\bibitem[H{\o}g et al.(2000)]{Hog2000}
H{\o}g, E. et al. 2000, \aap, 355, L27

\bibitem[Hurter et al.(2009)]{Hurter2009}
Hurter, C., Tissoires, B. \& Conversy, S. 2009, ``FromDaDy: Spreading Aircraft
Trajectories Across Views to Support Iterative Queries", IEEE Transactions on
Visualization and Computer Graphics, 15, 1017

\bibitem[Jordan et al.(2008)]{Jordan2008}
Jordan, D.D. et al. 2008, AAS 08-245, AAS/AIAA Space Flight Mechanics Meeting,
(AAS 08-245; Springfield; AAS), 499

\bibitem[MacDougall \& Henze(2003)]{Macdoughall2003}
MacDougall, P. \& Henze C. ``Fleshing-out pharmacophores with volume rendering
of molecular charge densities and hyperwall visualization technology,
226th ACS National Meeting New York, NY September 07-11, 2003.

\bibitem[Murman et al.(2004)]{Murman2004}
Murman, S.M., Aftosimis, M.J. \& Nemec, M. ``Automated Parameter Studies
Using a Cartesian Method'', AIAA 2004-5076 22nd AIAA Applied Aerodynamics
Conference, Providence, RI., 2004 

\bibitem[Perryman et al.(1997)]{ESA1997}
Perryman, M.A.C., et al., 1997, The Hipparcos and Tycho Catalogues, (SP-1200,
Noordwijk: ESA), 117

\bibitem[Sandstrom et al.(2003)]{Sandstrom2003}
Sandstrom, T.A., Henze, C., \& Levit, C. 2003,
Coordinated \& Multiple Views in Exploratory Visualization
(Piscataway: IEEE), 124

\bibitem[Stump et al.(2004)]{Stump2004}
Stump, G.M. et al. 2004, Aerospace Conference (Piscataway: IEEE), 6, 3885

\bibitem[SubbaRao et al.(2008)]{Subbarao08}
SubbaRao, M. U., Arag\'on-Calvo, M.A., Chen, H.W., Quashnock, J.M.,
Szalay, A.S. \& York, D.G. 2008, NJP, 10, 125015

\bibitem[Way, Gazis \& Scargle(2010)]{WGS2010}
Way, M.J., Gazis, P.R. \& Scargle, J.S. 2010, \apj, submitted
(arXiv:1009.0387v1)

\bibitem[Wong \& Bergeron(1997)]{Wong1997}
Wong, P.C. \& Bergeron, R.D. 1997, Scientific Visualization Overviews
Methodologies and Techniques (Washington: IEEE), 3

\bibitem[York et al.(2000)]{York2000}
York, D.G. et al. 2000, \aj, 120, 1579

\bibitem[Zhang et al.(2003)]{Zhang2003}
Zhang, J. et al. 2003, IEEE International Conference on Data Mining
(Washington: IEEE)

\end{thebibliography}
\end{document}